# Variational quantum algorithm for non-Markovian quantum dynamics

Peter L. Walters[a], Mohammad U. Sherazi[b], Fei Wang[a,c*]

[a] Department of Chemistry and Biochemistry, George Mason University, Fairfax, VA, US, 22033
[b] Department of Physics and Astronomy, George Mason University, Fairfax, VA, US, 22033
[c] Quantum Science and Engineering Center, George Mason University, Fairfax, VA, US, 22033

**ABSTRACT:** The simulation of non-Markovian quantum dynamics plays an important role in the understanding of charge and exciton dynamics in the condensed phase environment, and yet it remains computationally expensive on classical computers. We have developed a variational quantum algorithm that is capable of simulating non-Markovian quantum dynamics. The algorithm captures the non-Markovian effect by employing the Ehrenfect trajectories in the path integral formulation and the Monte Carlo sampling of the thermal distribution. We tested the algorithm with the spin-boson model on the quantum simulator and the results match well with the exact ones. The algorithm naturally fits into the parallel computing platform of the NISQ devices and is well suited for anharmonic system-bath interactions and multi-state systems.

## I.     Introduction

The simulation of quantum dynamics in the condensed phase environment can offer critical insight into the charge and exciton transfer processes in solutions, functional materials and biomolecules.[1–7] Particularly interesting and challenging is the simulation in the non-Markovian regime where the system's past trajectory influences its present state. In such case, the computation on the classical computers often scales exponentially with respect to the system size and the memory length. A quantum computer, on the other hand, can encode the exponential number of states with a linear number of qubits. Very recently, there has been a growing interest in the development of quantum algorithms for non-Markovian quantum dynamics.[8–12] Given the current stage of the quantum devices, it is unlikely that the full-scale quantum algorithms such as Shor's factoring[13] can be implemented in the near future. Therefore, a hybrid quantum-classical approach that utilizes variational quantum circuits[14,15] seems to be the practical and immediate application of near-term quantum computing. In this work, we present a variational quantum algorithm (VQA) for simulating non-Markovian quantum dynamics. The algorithm uses the ensemble-averaged Ehrenfest trajectories (EAET) to capture the non-Markovian effect and employs the "projected – Variational Quantum Dynamics" (p-VQD)[16] method to parametrize the circuit. In what follows, we will discuss the EAET formulation in Section II and the p-VQD algorithm in Section III. We will present the simulation results in Section IV and offer concluding remarks in Section V.

## II.    EAET formulation

We will use a quantum system linearly coupled to its harmonic bath as our model. The Hamiltonian of such can be written as

$$H_{tot} = H_0 + \sum_j \left[ \frac{P_j^2}{2m_j} + \frac{1}{2} m_j \omega_j^2 \left( x_j - \frac{c_j s}{m_j \omega_j^2} \right)^2 \right] \quad (1)$$

where $H_0$ is the system Hamiltonian, $s$ and $x_j$ denote the system and bath coordinates, respectively, and $c_j$ denotes the system-bath coupling. The strength weighted density of modes defines the spectral density,[17]

$$J(\omega) = \frac{\pi}{2} \sum_j \frac{c_j^2}{m_j \omega_j} \delta(\omega - \omega_j) \qquad (2)$$

The system is perturbed by a time-dependent driving force from the bath,[18]

$$H_s(t) = H_0 - \sum_j c_j\, s x_j(t) \qquad (3)$$

where

$$x_j(t) = x_{0,j} \cos \omega_j t + \frac{p_{0,j}}{m_j \omega_j} \sin \omega_j t + \frac{c_j}{m_j \omega_j} \int_0^t dt'\, s(t') \sin \omega_j (t - t') \qquad (4)$$

The non-local memory kernel in the last part of equation (4), termed the back-reaction[19] (i.e., kicking back by the system), is partially responsible for the non-Markovian effect. The other contribution is from the integration of the phase space variables $x_{0,j}$ and $p_{0,j}$ of the bath. The exponential scaling can be easily seen with discretized position and time. For instance, for a two-state system, the trajectory proliferation can be pictorially represented as in Figure 1,

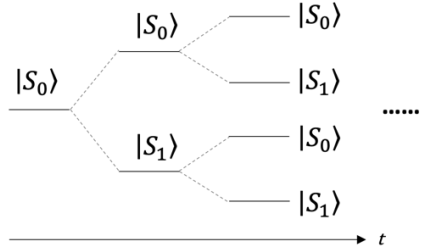

Figure 1. Exponential proliferation of trajectories.

Indeed, in the path integral formulation, every trajectory contributes to the dynamics, but with a different weight. In the Ehrenfest trajectory (ET) approximation, we replace all possible trajectories with one average trajectory with weighted positions. Specifically,

$$x_j(t) \cong x_{0,j} \cos \omega_j t + \frac{p_{0,j}}{m_j \omega_j} \sin \omega_j t + \frac{c_j}{m_j \omega_j} \int_0^t dt'\, \mathcal{Y}(t') \sin \omega_j (t - t') \qquad (5)$$

where $\mathcal{Y}(t')$ is the average trajectory of all possible positions at time $t'$. For example, for a two-state system in which the two localized states are also position eigenstates, $\mathcal{Y}(t') = p_0(t') \times s_0 + p_1(t') \times s_1$, where $p_0(t')$ and $p_1(t')$ are the populations of these two states at time $t'$, and $s_0$ and $s_1$ are the position values.

With the ET approximation, the dynamics can be solved by Markovian propagation,

$$i\hbar \frac{\partial}{\partial t} |\psi(t)\rangle = H_s(t)|\psi(t)\rangle \qquad (6)$$

To incorporate the thermal effect, Monte Carlo sampling of $x_0$ and $p_0$ from the Wigner distribution, equation (7), is performed.

$$W(\boldsymbol{x_0}, \boldsymbol{p_0}) = (\hbar\pi)^{-1} \prod_j \tanh\left(\frac{1}{2}\hbar\omega_j\beta\right) \exp\left[-\tanh\left(\frac{1}{2}\hbar\omega_j\beta\right)\left(\frac{m_j\omega_j x_{0,j}^2}{\hbar} + \frac{p_{0,j}^2}{m_j\omega_j\hbar}\right)\right] \quad (7)$$

The result is the ensemble averaged $|\psi(t)\rangle$, which we call the ensemble averaged Ehrenfest trajectory (EAET) approach. It draws inspirations from the ensemble averaged classical path (EACP) developed by Makri.[20] One appealing aspect of these approaches is that they are not limited to the harmonic bath linearly coupled to the system; the framework can be equally adapted to non-linear coupling and anharmonic environment,[19] provided its initial Wigner distribution is available.[21] We emphasize that the Wigner distribution automatically accounts for the zero-point energy effect of the bath, whereas the Boltzmann distribution does not.

Inserting equation (4) in the action integral and performing the integration over $\boldsymbol{x_0}$ and $\boldsymbol{p_0}$ with the Wigner distribution reproduce the Feynman-Vernon's influence functional $IF$,[22] in which

$$IF = Q \times R \quad (8)$$

The influence functional alters the system's free dynamics in the dissipative environment.

Here,

$$Q = \exp\left\{-\sum_j \frac{c_j^2}{2m_j\omega_j\hbar} \coth\left(\frac{1}{2}\hbar\omega_j\beta\right) \int_0^t dt' \int_0^{t'} dt'' \,\Delta s(t')\Delta s(t'') \cos\omega_j(t'-t'')\right\} \quad (9)$$

and

$$R = \exp\left\{\frac{i}{\hbar}\sum_j \frac{c_j^2}{m_j\omega_j} \int_0^t dt' \int_0^{t'} dt'' \,\Delta s(t')\bar{s}(t'') \sin\omega_j(t'-t'')\right\} \quad (10)$$

The $s^+$ and $s^-$ denote the forward and backward position, respectively, $\Delta s = s^+ - s^-$, and $\bar{s} = \frac{1}{2}(s^+ + s^-)$. Apparent from equation (8-9) is the non-Markovian effect induced by the double time integral. The non-Markovianity in $Q$ results from the integration of the phase space variables whereas the non-Markovianity in $R$ originates from the back-reaction term in equation (4). As pointed out by Makri, the temperature dependent $Q$ is related to the simulated emission and absorption of phonons, whereas $R$ the spontaneous emission.[23] In the EAET approximation, $Q$ remains unchanged and $R$ becomes

$$R \cong \exp\left\{\frac{i}{\hbar}\sum_j \frac{c_j^2}{m_j\omega_j} \int_0^t dt' \int_0^{t'} dt'' \,\Delta s(t')\mathcal{Y}(t'') \sin\omega_j(t'-t'')\right\} \quad (11)$$

Therefore, the EAET approximation preserves most of the non-Markovian effect by the double-time integral, as well as the quantum mechanical effect from the Wigner distribution and the back-reaction.

### III.    p-VQD algorithm

To propagate equation (6) on the quantum computer, we use the VQA approach. We adopt the p-VQD,[16] an optimization-based method, to construct the quantum circuit. First, define the loss function $L$ with the parametrized circuit $C(\boldsymbol{\theta})$ as

$$L(d\boldsymbol{\theta}, dt) = \frac{1 - |\langle 0|C^\dagger(\boldsymbol{\theta})e^{iHdt}C(\boldsymbol{\theta} + d\boldsymbol{\theta})|0\rangle|^2}{dt^2} \tag{12}$$

in which the factor $dt^2$ in the denominator is to ensure that $L$ is independent of the time step size. If the parametrized circuit is comprised of generalized Pauli operators, then the gradient can be computed exactly using the parameter shift rule,[24,25]

$$\frac{\partial}{\partial d\theta_i} L(d\boldsymbol{\theta}, dt) = \frac{1}{2}\left\{L\left(d\boldsymbol{\theta} + \frac{\pi}{2}e_i, dt\right) - L\left(d\boldsymbol{\theta} - \frac{\pi}{2}e_i, dt\right)\right\} \tag{13}$$

where $e_i$ is the vector in the $i$-th direction. In this work, we choose to use the ADAM stochastic algorithm for the optimization.[26] The advantages of the p-VQD compared to the time-dependent variational algorithm[27] are that it circumvents the numerical instability arising from matrix inversion, it scales linearly with the number of parameters, and it avoids the barren plateaus.

## IV. Simulation results

In the following, we use spin-boson model to test the algorithm. The Hamiltonian in the EAET limit can be written as

$$H(t) = \hbar\Omega\sigma_x - \left(\sum_j c_j x_j(t)\right)\sigma_z \tag{14}$$

in which $x_j(t)$ is given by equation (5). We choose the bath to have the Ohmic spectral density

$$J(\omega) = \frac{\pi}{2}\hbar\xi\omega e^{-\omega/\omega_c} \tag{15}$$

where the dimensionless $\xi$ is the Kondo parameter that determines the strength of the system-bath coupling, and $\omega_c$ is the cutoff frequency. We use 60 oscillators of different frequencies in the numerical calculation, following the discretization procedure given by Walters et al.[28] For a two-level system which requires one qubit, there exits an exact ansatz for the unitary operation that employs the $ZXZ$ decomposition[29]

$$C(\boldsymbol{\theta}) = e^{i\theta_1}R_z(\theta_2)R_x(\theta_3)R_z(\theta_4) \tag{16}$$

We show the simulation results in Figure 1-4. Specifically, Figures 1 and 2 show the population dynamics with parameters $\xi = 1.2$, $\omega_c = 2.5$, $\beta = 0.2$, and $\Omega = 1$. The system is initially in the reactant state. In Figure 1, the "Exact" result is from the numerically exact QuAPI calculation,[30] the "Ehrenfest" result is obtained by numerically solving the differential equation (6) with the RK4 method and then averaging over the Monte Carlo points (i.e. EAET), the "p-VQD" result is obtained from numerically solving equation (6) with the parametrized circuit in equation (16) and the ADAM optimization, and the "Simulator" result is obtained by using the p-VQD approach and the measurement of the circuit with 50,000 shots. In Figure 2 (a)-(d), the "Ehrenfest 1 IC" is the simulation result of the ET approximation with one initial condition, $\boldsymbol{x}_0$ and $\boldsymbol{p}_0$, randomly chosen. The "Ehrenfest 10,000 ICs" is from the EAET, averaging over 10,000 initial conditions. Similarly, Figure 2 and 3 use parameters $\xi = 0.3$, $\omega_c = 5$, $\beta = 5$, and $\Omega = 1$.

In Figures 1 and 3, the EAET matches well with the exact benchmark. It captures the correct timescale and the dynamical behavior. The small discrepancy comes from the fact that the ET ignores some quantum interference effect from the different possible trajectories, and therefore tends to decay more. But due to the decoherence effect of the bath, the dynamical eventually becomes more classical that follows the Eherenfest trajectory and the EAET results are still in quantitative agreement with the exact ones. In Figures 2 and 4, the "Simulator" result eventually deviates from the "p-VQD" result due to the shot noise, however, in the EAET simulation, they match perfectly. Therefore, by the ensemble averaging, the shot error can be reduced due to its random nature. Figures 2 and 4 also show that mild Monte Carlo points such as 10,000 will make the results converge well.

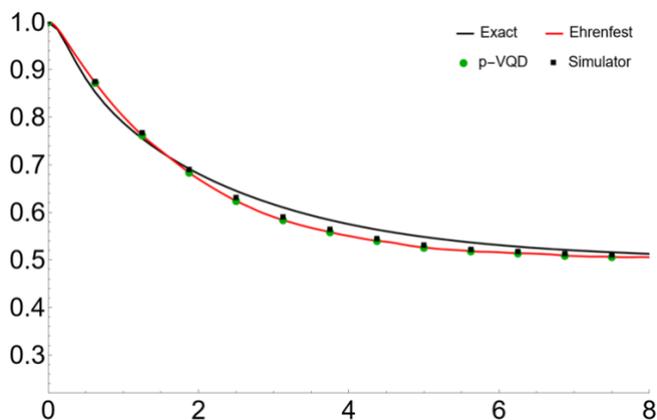

Figure 1. Population dynamics for a symmetric two-level system coupled to a harmonic bath with parameters $\Omega = 1, \xi = 1.2, \omega_c = 2.5, \beta = 0.2$ and the system initially populated in the reactant state.

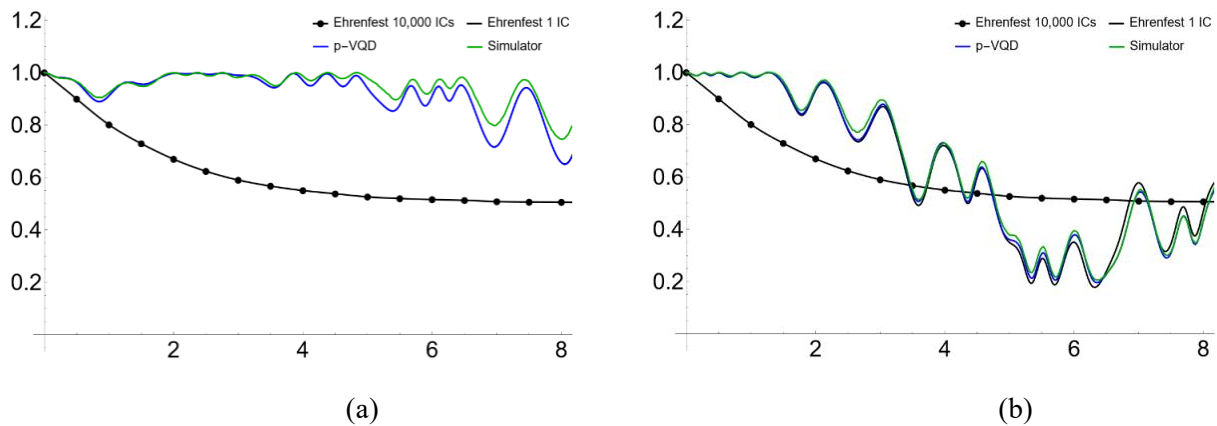

(a)                                              (b)

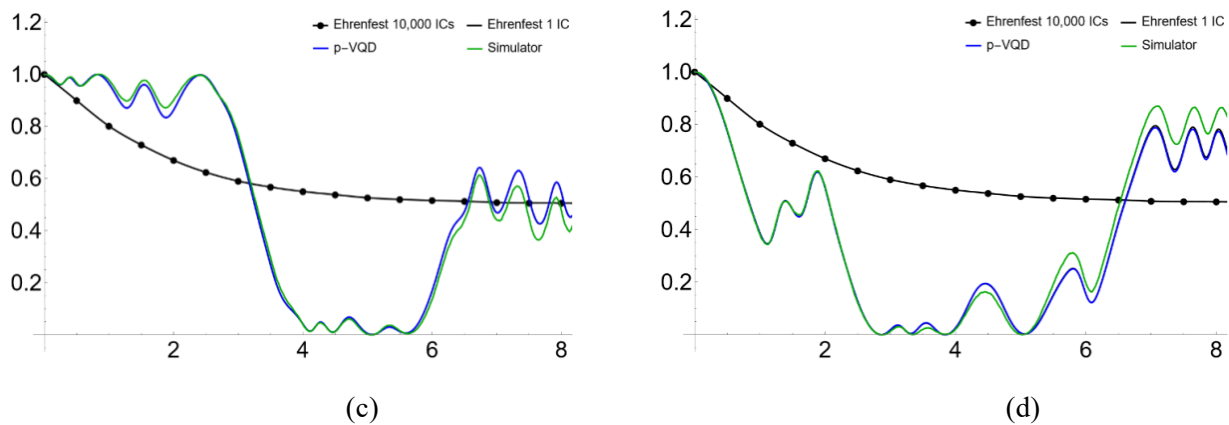

(c)　　　　　　　　　　　　　　　　(d)

Figure 2 (a)-(d). Population dynamics for a symmetric two-level system coupled to a harmonic bath with one initial condition and parameters $\Omega = 1$, $\xi = 1.2$, $\omega_c = 2.5$. The system is initially in the reactant state.

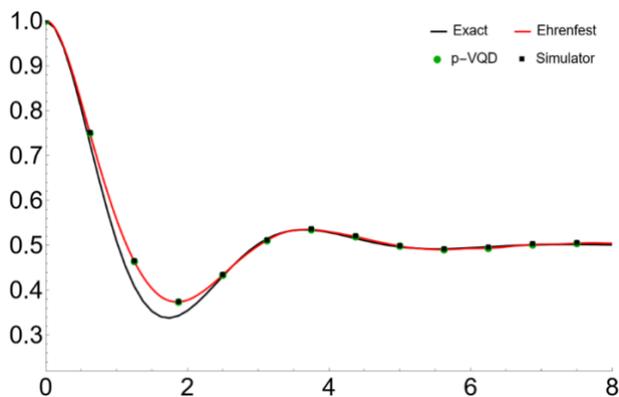

Figure 3. Population dynamics for a symmetric two-level system coupled to a harmonic bath with parameters $\Omega = 1$, $\xi = 0.3$, $\omega_c = 5$, $\beta = 5$ and the system initially populated in the reactant state.

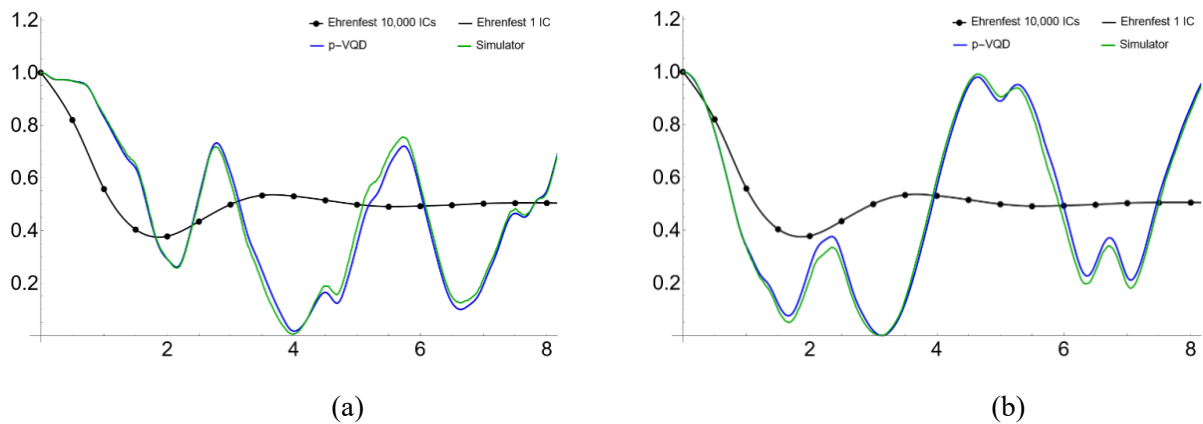

(a)　　　　　　　　　　　　　　　　(b)

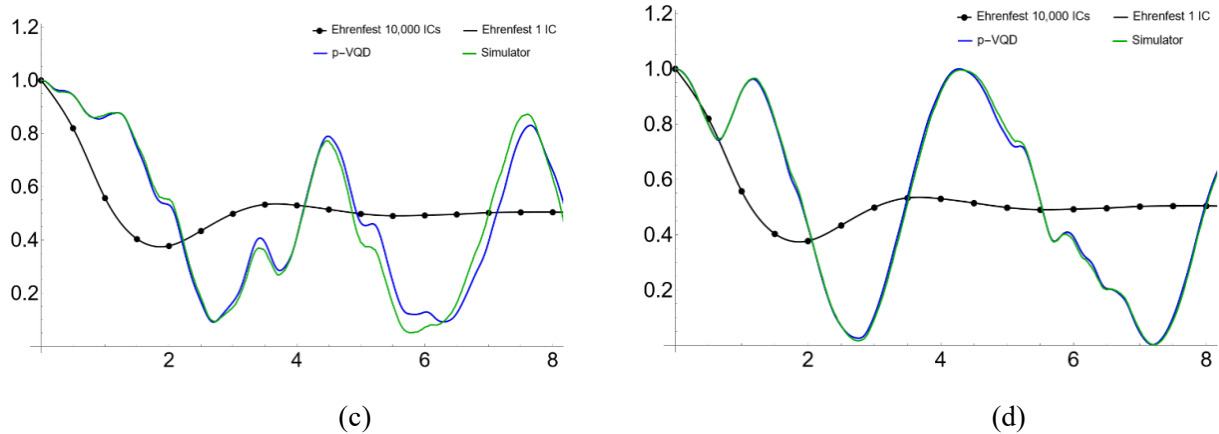

(c)                      (d)

Figure 4 (a)-(d). Population dynamics for a symmetric two-level system coupled to a harmonic bath with one initial condition and parameters $\Omega = 1$, $\xi = 0.3$, $\omega_c = 5$. The system is initially in the reactant state.

## V. Conclusion

In summary, we have developed a variational quantum algorithm that is able to simulate non-Markovian quantum dynamics at finite temperature. The algorithm is numerically stable and takes advantage of both the classical and quantum computing. The Monte Carlo sampling can be performed efficiently on a classical computer and the wavefunction overlap evaluated in equation (12) can be handled with a linear number of qubits on a quantum computer. In addition, since each Ehrenfest trajectory originated from the Monte Carlo points can be propagated independently, the EAET algorithm can be implemented parallelly on the NISQ devices. Furthermore, the algorithm can be well adapted to multi-state problems,[31,32] anharmonic bath and non-linear system-bath coupling.[19]

## AUTHOR INFORMATION

### Corresponding Author

Fei Wang — Department of Chemistry and Biochemistry, Quantum Science and Engineering Center, George Mason University, Fairfax, VA, US, 22033

### Authors

Peter L. Walters — Department of Chemistry and Biochemistry, George Mason University, Fairfax, VA, US, 22033

Mohammad U. Sherazi — Department of Physics and Astronomy, George Mason University, Fairfax, VA, US, 22033

### Notes

The authors declare no competing financial interest.

## ACKNOWLEDGEMENTS

This work is supported by the National Science Foundation (NSF) under the Award 2320328, and the George Mason University's startup fund. This work used the SDSC Expanse CPU of the Explore ACCESS through allocation CHE220009 and CHE240132 from the Advanced Cyberinfrastructure Coordination Ecosystem: Services & Support (ACCESS) program, which is supported by National Science Foundation grants 2138259, 2138286, 2138307, 2137603, and 2138296.